\documentclass[twocolumn]{aastex701}
\usepackage{amsmath}

\usepackage{tabularray}
\usepackage{graphicx}

\newcommand{\cyanade}{$\mathrm{sNH_4^+CN^-}$}
\newcommand{\cyanate}{$\mathrm{sNH_4^+OCN^-}$}
\newcommand{\formate}{$\mathrm{sNH_4^+HCOO^-}$}
\newcommand{\hydrosulfide}{$\mathrm{sNH_4^+SH^-}$}
\newcommand{\acetate}{$\mathrm{sNH_4^+CH_3COO^-}$}
\newcommand{\carbamate}{$\mathrm{sNH_4^+NH_2COO^-}$}

\begin{document}

\title{Ammonium salt formation and abundance in protoplanetary disks}

\author[orcid=0000-0003-0522-5789,sname='Ruaud']{Maxime Ruaud}
\affiliation{3 Boussac, Augan, France}
\email[show]{maxime.ruaud@gmail.com}  

\author[orcid=0000-0001-8063-8685,sname='Loison']{Jean-Christophe Loison}
\affiliation{Université Bordeaux, CNRS, Bordeaux INP, ISM, UMR 5255, 33400 Talence, France}
\email{} 

\author[orcid=0000-0000-0000-0001,sname='Gorti']{Uma Gorti}
\affiliation{Carl Sagan Center, SETI Institute, Mountain View, CA, USA}
\affiliation{NASA Ames Research Center, Moffett Field, CA 94035, USA}
\email{} 

\begin{abstract}
Ammonium salts may represent an important reservoir of volatile species in Solar system primitive bodies, but the question of how and when these salts can form during the star formation process remains unknown. In this paper, we use thermo-chemical models to study the formation of ammonium salts during the protoplanetary disk stage. We show that ammonium salts form efficiently in the inner disk midplane (i.e. $r \lesssim 50 $ au), inside the comet forming region. In this region, our model predicts that almost all the available nitrogen is in the form of salts  (i.e. mainly in ammonium cyanate) at the surface of grains after evolving for 10 Myrs. For sulfur, we show that almost all the available S is in the form of ammonium hydrosulfide in the inner disk midplane. We show that inside  $r\sim 30$ au, ammonium salt formation is enhanced by a cosmic-ray-driven sink effect that progressively converts gas-phase CO and N$_2$ into carbon dioxide and salts, respectively, at the surface of grains on a timescale $\gtrsim 1$ Myr. This impacts the location of the CO and N$_2$ radial snowlines which both shift closer to the star as a function of time.

\end{abstract}


\section{Introduction} 

Nitrogen is one of the most abundant elements in the universe. As an essential component of prebiotic molecules like amino acids and nucleobases, nitrogen is critical to astrobiology. In the interstellar medium, in protoplanetary disks, and in the cold outer solar system, nitrogen is mostly found as a gas, N$_2$, or frozen in ice as N$_2$ or in the form of volatile species such as NH$_3$ and HCN. The extreme volatility of nitrogen makes it a sensitive tracer of disk conditions such as the radial temperature gradient, and processes such as material transport and loss of volatiles as a disk evolves to form planetary systems. 

Key constraints on the nitrogen reservoirs in disks and the extent to which it is removed with the gas come from our own solar system. Nitrogen is preserved to some extent on cold, massive objects in the outer solar system (e.g., Pluto, Titan, Triton), but is believed to be missing in comets and heavily depleted on Earth and asteroids in the inner solar system \citep[e.g.,][]{Bergin15}. The low nitrogen content can be interpreted in two ways: (1) either nitrogen is truly depleted  due to a combination of inefficient trapping of volatile nitrogen at formation locations and/or subsequent evolutionary loss, or (2) a significant fraction of the nitrogen is locked up in a more refractory form. 

Recent data suggest that ammonium (NH$_4^+$) salts may constitute a hidden refractory reservoir of nitrogen. The Rosetta mission to Comet 67P found strong evidence that a significant fraction of the missing nitrogen is locked in relatively less volatile ammonium  salts embedded in cometary dust grains. 
Ammonium salts have also been inferred on other comets \citep{Munaretto2026} and icy bodies like Ceres \citep{DeSanctis24}.

Salts are also suspected to be present in asteroids \citep{Rivkin22,Naraoka23,Laize24,Glavin25}. The presence of ammoniated salts in  ices of asteroidal parent bodies could have facilitated melting and, therefore, the formation of phyllosilicates, as well as the organic material seen today in asteroids \citep{Rivkin22}. Samples returned from Bennu by the OSIRIS-Rex mission indeed show the presence of abundant N-rich molecules likely formed by aqueous alteration in ammoniated fluids \citep{Glavin25,Sandford25}.

Whereas asteroid materials have undergone significant alteration subsequent to their formation history, cometary materials maintain a more pristine record of disk conditions and the available nitrogen reservoir during solar system formation. 

In comets, the main nitrogen carriers NH$_3$ and HCN have abundances  relative to water $\lesssim 1$\%. In the sun, the N/O ratio is  $\sim 11\%$. Comets have nearly solar oxygen abundances, and therefore the N/O ratio is a factor $\sim 10$ lower in the volatile components. Estimates made in comet 67P from enhanced ammonia degassing during a late stage dust impact suggest that 90\% of the observed NH$_3$ during this event was a sublimation product from evaporating salts. This could explain most of the nitrogen deficiency in this comet \citep{Altwegg20}.  \citet{Poch20}, based on the analysis of the reflectance spectrum obtained of comet 67P, estimate that $\sim 50\%$ of the overall nitrogen reservoir (including refractory material) could be in ammonium salts. A large fraction of sulfur, for which the distribution has been a longstanding puzzle in the ISM, is also suspected to be in the form of salts (i.e. \hydrosulfide) at surface of 67P \citep{Altwegg22}.

Recent JWST observations reveal the presence of OCN$^-$, likely associated with ammonium cyanate (NH$_4^+$OCN$^-$), in the ices of protoplanetary disks \citep{Sturm23,Potapov25, Bergner2026} and provide a more direct motivation to investigate the formation of salts in protoplanetary disks.  These JWST data present indirect evidence that ammonium salts could form in icy disk midplanes. \citet{Potapov25} also report the detection of NH$_4^+$ and ammonium carbamate in the ice of the disk d216-0939.

Laboratory experiments on interstellar ice analogs, as well as calculations, show that ammonium salts readily form at cryogenic temperatures (i.e., at temperatures as low as 10 K) and with an efficiency that depends only slightly on the H$_2$O dilution \citep{vanBroekhuizen04}; i.e. provided their constituent species are abundant in the ice. The formation of these salts proceed through acid-base reactions in the ice. When desorbing in the gas-phase, these salts dissociate mainly into stable products NH$_3$ + HX \citep{Kruczkiewicz21,Loison25}. This implies that the apparent desorption temperature of certain species will be higher if they are in salt form. Salts could moreover, be a repository of other elements believed to be depleted in disks and in the solar system. Notably, substantial amounts of sulfur-bearing species have also been detected in comet 67P \citep{Altwegg22} and in the atmospheres of giant exoplanets (e.g., SO$_2$ and NH$_3$ were detected in WASP-107b, \citet{Welbanks24}), suggesting that sulfur may also be preserved as salts in the disk \citep{Nakazawa26}.

In this paper we study the formation of salts in protoplanetary disks and show that ammonium salts can form efficiently in the inner disk midplane at comet forming radii. This paper is organized as follows. In Section \ref{sec:model_description} we describe our modeling framework. In Section \ref{sec:results} we present our modeling results. The formation and abundance of ammonium salts in the midplane of protoplanetary disks is described in \S \ref{sec:salt_form} and \S \ref{sec:sink_N2}. In \S \ref{sec:sulfur} we study the effects of varying the elemental abundance of sulfur on the results and discuss its impact on the midplane partitioning of sulfur (\S \ref{sec:sulfur}) and nitrogen (\S \ref{sec:nitrogen}). In Section \ref{sec:discussion} we discuss our results in the context of comets. Section \ref{sec:conclusion} summarizes our work.

\section{Model description} 
\label{sec:model_description}

\subsection{Disk model}
We use the disk modeling framework presented in \citet{Ruaud19}. We provide a brief summary of the modeling procedure here. 
Given a minimal set of physical parameters (i.e., stellar properties and the gas and dust surface density distributions), the disk physical structure is determined by solving for vertical hydrostatic equilibrium, coupled iteratively with gas heating and cooling, as well as steady state chemistry. Dust grains follow an MRN type distribution ($n_d(a) \propto a^{-3.5}$) with $a_\mathrm{min} = 0.005$ $\mu$m and $a_\mathrm{max} = 1$ cm and are composed of silicates and carbon. Dust grains settle vertically with a scale height set by the gas phase turbulence (we assume $\alpha=5\times 10^{-3}$). The resulting structure is then used to compute the time-dependent gas grain chemistry. For grain chemistry, the ice surface and ice mantle are treated as two separate phases in interaction, and grain surface chemistry includes the diffusion of the chemical species on the surface, two-body reactions, photo-reactions, and thermal and nonthermal desorption mechanisms; i.e. photodesorption and chemical desorption. The photodesorption yield is assumed to be similar for all species and equal to $10^{-3}$ molecules per photon \citep{Ruaud24}. For thermal diffusion, diffusion energy barriers ($E_\mathrm{diff}$) are poorly constrained (see discussion in \citet{Ruaud19} and \citet{Furuya22}). As a consequence, $E_\mathrm{diff}$ for each species are computed assuming a fixed fraction of its binding energy ($E_\mathrm{bind}$). We assume $E_\mathrm{diff}/E_\mathrm{bind} = 0.4$ for surface species and $E_\mathrm{diff}/E_\mathrm{bind} = 0.8$ for species in the mantle. In the following, molecules denoted as sX refer to molecules present in the ice and include both surface and mantle abundances.

\subsection{Ammonium salts chemistry}

\begin{table*}
    \centering
    \begin{tabular}{llc}
        \hline 
        \hline 
        Reactants &  Product & Product name \\
        \hline 
         sNH$_3$ + sHCN        & sNH$_4^+$CN$^-$       & Ammonium cyanide \\
         sNH$_3$ + sHNCO        & sNH$_4^+$OCN$^-$      & Ammonium cyanate \\
         sO +  sNH$_4^+$CN$^-$ &  sNH$_4^+$OCN$^-$    \\
         sNH$_3$ + sHCOOH      & sNH$_4^+$HCOO$^-$     & Ammonium formate \\
         sNH$_3$ + sH$_2$S         & sNH$_4^+$SH$^-$       & Ammonium hydrosulfide \\
         sNH$_3$ + sCH$_3$COOH & sNH$_4^+$CH$_3$COO$^-$& Ammonium acetate \\
         sNH$_3$ + sNH$_2$COOH & sNH$_4^+$NH$_2$COO$^-$& Ammonium carbamate \\
         \hline 
    \end{tabular}
    \caption{Ammonium salts formation route considered in this study.}
    \label{tab:tab1}
\end{table*}

\begin{table}
    \centering
    \begin{tabular}{lccc}
        \hline 
        \hline 
        Species & A ($s^{-1}$) & $E_\mathrm{bind}$ $(K)$ & Refs. \\
        \hline 
        \cyanade & $1.6\times 10^{17}$ & 7417 & (1) \\
         \cyanate & $2.9\times 10^{17}$ & 8117 & (1) \\
         \formate & $2\times 10^{18}$ & 9426 & (1) \\
         \hydrosulfide & $1\times 10^{12}$ & 6000 & (2)  \\
         \acetate & $1.3\times 10^{19}$ & 9408 & (1) \\
         \carbamate &$6.9\times 10^{19}$ & 12166 & (1) \\
         \hline 
    \end{tabular}
    \caption{Thermal desorption parameters used in the model. A is the pre-exponential factor and $E_\mathrm{bind}$ the binding energy. The rate coefficients for thermal desorption are given by $A \times \exp{(-E_\mathrm{bind} / T_\mathrm{d})}$.  References: (1) \citet{Ligterink23}, (2) \citet{Vitorino24}.}
    \label{tab:tab2}
\end{table}

We introduce salt formation by reaction of sNH$_3$ with sHCN, sHNC, sH$_2$S, sHNCO, sHCOOH, sCH$_3$COOH and sNH$_2$COOH, which are all abundant in the icy midplane of disks \citep{Ruaud19}. Our earlier chemical network did not include sNH$_2$COOH and we have now added this species with new formation and destruction pathways (see Table \ref{tab:tab3}). Tables \ref{tab:tab1}  and \ref{tab:tab2} summarize salt formation pathways considered in this study, and the desorption parameters used for each of these salts. Salts form due to proton transfer from an acid to a base, and this reaction is exothermic \citep{Loison25}. Therefore, ammonium salts do not exist in the gas phase; they are only stable when the resulting ions from the proton transfer can be sufficiently stabilized by interactions with water molecules to compensate for the exothermicity. We thus assume that when salts desorb, they dissociate into the stable products NH$_3$ + HX. We also note that we do not account for any destruction processes in the ice other than desorption. As a result, our results should be regarded as upper limits on the abundance of ammonium salts in protoplanetary disks.

\section{Results}
\label{sec:results}
In what follows, we consider a fiducial disk model with a total gas mass $M_\mathrm{g} =10^{-2} M_\odot$ and a dust to gas mass ratio $\Sigma_\mathrm{d} / \Sigma_\mathrm{g}= 10^{-2}$. The initial chemical conditions in the disk are obtained by evolving a dark cloud model for 1 Myr with constant  physical parameters (i.e. $n_\mathrm{H} = 2\times 10^4$ cm$^{-3}$, $T_\mathrm{g} = T_\mathrm{d} = 10$ K and $A_V = 10$ mag). Initial elemental abundances are given in Table \ref{tab:elem_abun} and correspond to abundances derived in the $\zeta $ Oph diffuse cloud \citep{Savage96} except for S for which we study the effect of varying its initial elemental abundance. Understanding how sulfur is distributed in the interstellar medium has been a persistent challenge. In the densest parts of the ISM, large quantities of sulfur are unaccounted for compared to its observed abundance in diffuse clouds. However, recent detection of substantial amounts of H$_2$S in comet 67P point to an abundant reservoir of sulfur possibly in the form of salts \citep{Altwegg22}. In the fiducial model, we use a low elemental abundance of S of $10^{-6}$, consistent with depleted values obtained in the dense ISM and in disks \citep[e.g.][]{Jenkins09,Kama19}. By adopting this depleted elemental abundance, we implicitly assume that the missing sulfur is locked in an unidentified refractory or semi-refractory reservoir. We study the effect of increasing its elemental abundance to its Solar value in \S \ref{sec:sulfur}. 

We note that even though ammonium salt chemistry is taken into account in the dark cloud model, their computed abundances are low. This could be due to the fact that our model assumes that reactions occur only after species diffuse across grain surfaces. The low temperatures in the cloud environment suppress thermal diffusion of molecules with significant diffusion energy barriers within the ice, thereby limiting their reactivity. This diffusion-limited treatment likely underestimates salt formation at low temperatures, as it neglects non-diffusive reaction mechanisms that may operate in interstellar ices \citep{Jin20}. Therefore, abundances of salts could be higher than computed by our disk models if the initial abundances from the dark cloud are higher, as indicated by  recent JWST observations in dense clouds where NH$_4^+$ and OCN$^-$ have been inferred \citep{McClure23}.

Nitrogen in the disk is initially all on ices and is partitioned into the following species: 48\% sN$_2$, 20\% sCH$_3$NH$_2$, 13\% sHCN, 6\% sNH$_3$ and $\sim 10\%$ in other ice chemical species including sNO, sHNO and sHC$_3$N. Using this initial disk composition, the chemistry in the disk is then solved for each $(r,z)$ and for $t=10$ Myr.

\begin{table}
    \centering
    \begin{tabular}{l c}
    \hline
    \hline
    Element 	& Abundance \\
    \hline
    He      & 0.1       \\
    H		& 1.0 	   \\
    C       & 1.4 $\times 10^{-4}$  \\
    O       & 3.2 $\times 10^{-4}$ \\
    N		& 6.2$\times 10^{-5}$ \\
    Si      & 1.7$\times 10^{-6}$  \\
    Mg 		& 1.1$\times 10^{-6}$ \\
    S       & 1.0 $\times 10^{-6}$ $-$  1.5 $\times 10^{-5}$\\
    Fe		& 1.7 $\times 10^{-7}$   \\
    PAH     & 1.0 $\times 10^{-9}$ \\
    \hline
    \end{tabular}
    \caption{Gas-phase elemental abundances used in the chemical modeling.}
    \label{tab:elem_abun}
\end{table}

\subsection{Salts formation and abundance}
\label{sec:salt_form}

\begin{figure*}
    \centering
    \includegraphics[width=1\linewidth]{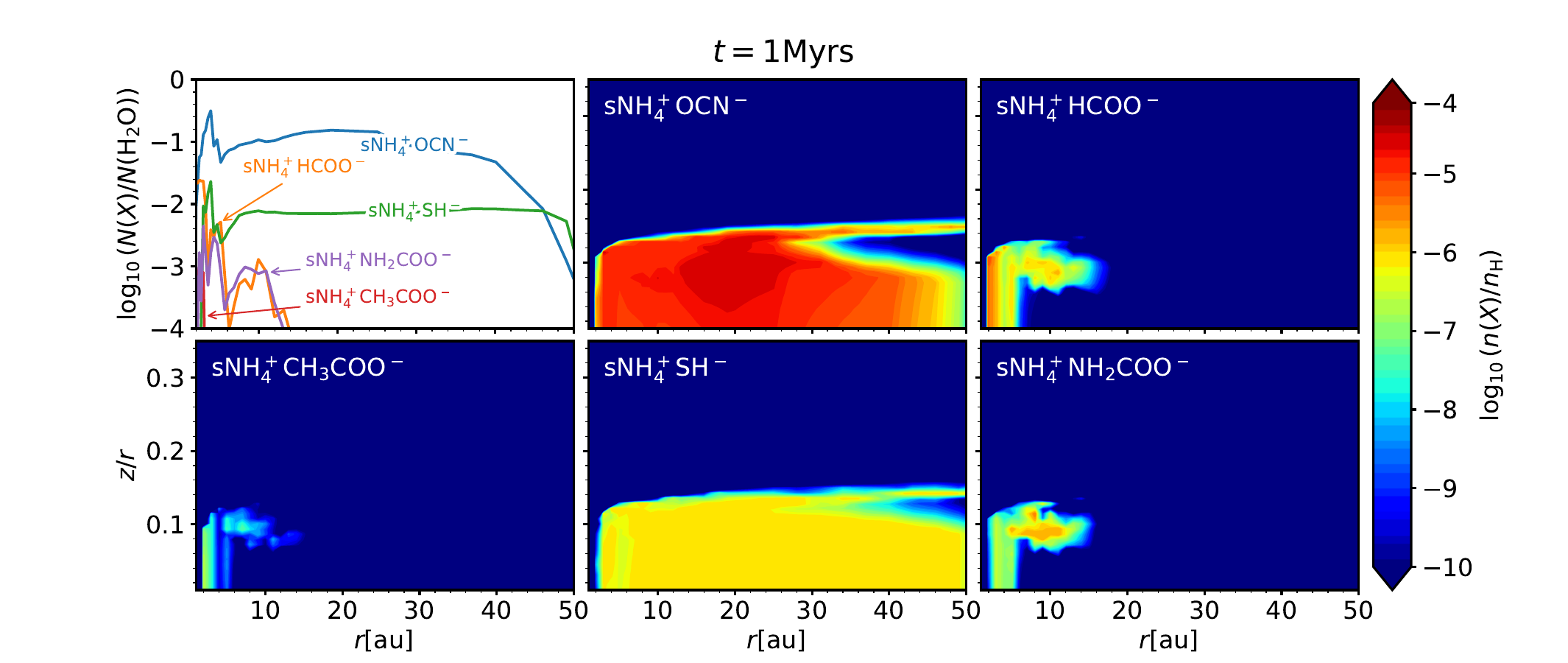}\\
    \includegraphics[width=1\linewidth]{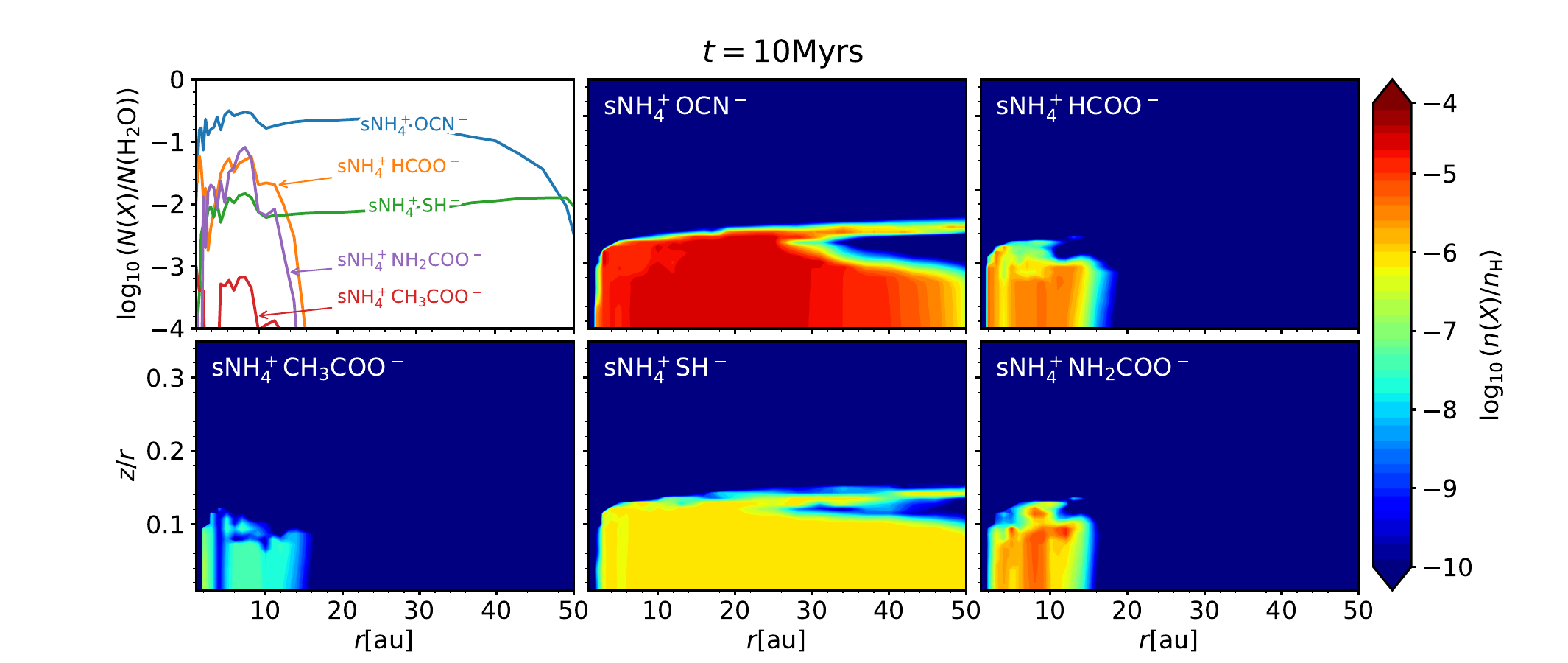}
    \caption{Computed abundance maps of ammonium cyanate, ammonium formate, ammonium acetate, ammonium hydrosulfide and ammonium carbamate in the solid phase for $r < 50$ au and computed after 1 Myr and 10 Myrs. We also show the ratio of the vertically integrated abundances of these salts relative to water ice.}
    \label{fig:fig1}
\end{figure*}

Figure \ref{fig:fig1} presents the abundance maps of ammonium cyanate (\cyanate), ammonium formate (\formate), ammonium hydrosulfide (\hydrosulfide), ammonium acetate (\acetate) and ammonium carbamate (\carbamate) in the solid phase within the inner disk ($r < 50$ au) and computed after 1 Myr (top) and 10 Myrs (bottom). The abundance of ammonium cyanide (\cyanade) is not shown here because of its very low abundance (typically lower than $10^{-10}$ relative to $n_\mathrm{H}$ throughout the disk). The figure also shows the ratio of the vertically integrated column densities of these salts relative to water ice. 

As can be seen, the efficiency with which salts can form is sensitive to the radial location within the disk. In our chemical model, the formation of ammonium salts is through reactions on grain surfaces and therefore depends on the availability and mobility of their constituent species on icy surfaces. In \citet{Ruaud19}, we show that species like sNH$_3$, sHCN, and sH$_2$S are abundant throughout the disk midplane, while species like sHNCO, sHCOOH, and sCH$_3$COOH preferentially form in the inner disk midplane, where higher temperatures favor their formation (due to an increased mobility in the ice). 
sNH$_3$ is relatively immobile on icy grain surfaces due to its high binding energy to water ice (approximately $5500$ K). Consequently, salt formation is primarily controlled by the mobility of acidic reactants like sHCN, sHNCO and sH$_2$S which is efficient in the inner disk midplane (i.e. $r \lesssim 50$ au). Outside $r\sim 50$ au, grains are too cold for these species to efficiently diffuse on icy grain surfaces and salt formation is therefore inefficient. In the following, we only consider regions in the disk with $r \lesssim 50$ au, where ammonium salts efficiently from. This region coincides with the comet forming region and is therefore relevant to ices located inside the Kuiper belt for our solar system. 

We find that ammonium cyanate (\cyanate) is the most abundant salt in our fiducial model inside $r\sim 40$ au, with a formation pathway 
\[
   \mathrm{sNH_3  + sHCN} \rightarrow \mathrm{sNH_4^+CN^-}
\]
followed by
\[
   \mathrm{sNH_4^+CN^-} \xrightarrow{\mathrm{sO}} \mathrm{sNH_4^+OCN^-}
\] 
These reactions are driven by the relative mobility of sHCN on icy surfaces. In the first step, ammonium cyanide can form efficiently by the reaction of sNH$_3$ with sHCN inside $r\sim 40$ au (which are both abundant at $t=0$ of the disk model). There is no theoretical or experimental study of the second step, but we argue that it is reasonable to assume that this reaction will be rapid on ices due to the following: 
\begin{enumerate}
\item[(i)] There is no theoretical or experimental study of the  O + CN$^-$  reaction, but in the gas phase, CN$^-$ reacts rapidly with H\citep{Lochmann2024}. Moreover, anions in general also react rapidly with oxygen atoms \citep{Eichelberger2007,Yang2011}. Therefore, we expect that CN$^-$ will react rapidly with O atoms to form OCN$^-$. 
\item[(ii)]The OCN$^-$ anion is expected to have an interaction energy with H$_2$O molecules in the ice that is quite similar to that of CN$^-$, i.e. this energy mainly depends on the size of the anions \citep[see][]{Zhu98}, which is similar in both cases. The exothermicity of the sO + \cyanade $\rightarrow$ \cyanate~ is therfore close to the exothermicity of the O + CN$^-$ $\rightarrow$ OCN$^-$ reaction (equal to 5.58 eV). We can reasonably assume that the O + CN$^-$ reaction in the ice and in the gas phase will be similar and fast.
\end{enumerate}

As seen in Fig \ref{fig:fig1}, \cyanate~ is present  throughout the disk midplane (i.e. at $z/r \lesssim 0.1 - 0.15$) with abundances greater than few $10^{-5}$. A small fraction of \cyanate~ also forms directly from sNH$_3$ + sHNCO.

Ammonium hydrosulfide (\hydrosulfide) is the second most abundant salt in our model outside $r\sim 15$ au (i.e. after ammonium cyanate). It 
efficiently forms through the reaction sNH$_3$ + sH$_2$S and is present inside $r \sim 50$ au with a relatively constant abundance of $\sim 10^{-6}$ with respect to $n_\mathrm{H}$. We note that the initial elemental abundance of sulfur in our fiducial model is $10^{-6}$ and therefore, most of the sulfur inside $r \sim 50$ au is in the form of salts. Later, in \S \ref{sec:sulfur}, we explore a model where we increase the initial sulfur abundance.

sHCOOH, sCH$_3$COOH and sNH$_2$COOH have diffusion energy barriers comparable to or higher than that of NH$_3$. As a consequence, ammonium formate (\formate), ammonium acetate (\acetate) and ammonium carbamate (\carbamate) form only at radii $r < 15$~au, where the temperature is high enough to have an efficient diffusion of their constituent species at the surface of grains.

\subsection{Sink effect on gas-phase N$_2$}
\label{sec:sink_N2}
Fig. \ref{fig:fig1} shows that the abundance of ammonium salts increases from $t=1$ Myr to $t=10$ Myrs in the inner disk midplane. This increase is explained by a sink effect that converts gas-phase N$_2$ into sNH$_3$ and subsequently to ammonium salts after $t=1$ Myr.

This sink effect on timescales $\gtrsim 1$ Myr, is driven by cosmic rays, for which we use a depth dependent expression from \citet{Padovani18} \citep[see also][]{Ruaud19}. $\mathrm{H_3^+}$ is produced by reaction of H$_2$ with cosmic rays\footnote{We include X-ray ionization in the model, but in the disk midplane ($z/r \lesssim 0.15$) formation of $\mathrm{H_3^+}$ is dominated by cosmic rays. In the disk midplane, the cosmic ray ionization rate computed from the depth dependent expression of \citet{Padovani18} ranges between $\zeta_\mathrm{CR}\sim 10^{-17}$ s$^{-1}$ in the inner regions and $\zeta_\mathrm{CR}\sim 10^{-16.5}$ s$^{-1}$ in the outer regions, where the gas column density is lower.} that forms H$_2^+$, which subsequently reacts with H$_2$ to form H$_3^+$.  Gas-phase N$_2$ then reacts with H$_3^+$ to form N$_2$H$^+$ and H$_2$.  When CO is present in the gas-phase, most of N$_2$H$^+$ reacts with CO to reform N$_2$ and HCO$^+$. However, when CO is absent from the gas-phase, N$_2$H$^+$ mainly recombines with electrons to mainly reform N$_2$ (95\%) as well as NH (5\%). NH can then accrete at the surface of grains to form sNH$_3$ by successive hydrogenation.

The efficiency of this process is therefore conditioned by the absence of CO from the gas-phase. In \citet{Ruaud19}, we show that in the shielded regions of the inner disk midplane, CO is converted to sCO$_2$ at the surface of the ice by a process that is also driven by cosmic rays. In this process, cosmic ray-generated UV photons dissociate water ice to form sOH which can then react with sCO to form sCO$_2$. The photodissociation timescale of water ice by this process is around 1 Myr for a typical cosmic ray flux of $10^{-17}$ s$^{-1}$ \citep[see][]{Ruaud19}. This conversion of CO into sCO$_2$ as well as the sink effect described for N$_2$ was also found by \citet{Furuya14} and \citet{Aikawa15}. \citet{Furuya14} also included effects of vertical mixing  and find that the N$_2$ sink effect may be reduced with strong levels of turbulence ($\alpha=10^{-2}$). However, this is higher than typical values ($\alpha \sim 10^{-3-4}$) in disks \citep[e.g.][]{Flaherty18,Flaherty20} and we therefore do not consider vertical mixing here.

\begin{figure}
    \centering
    \includegraphics[width=1\linewidth]{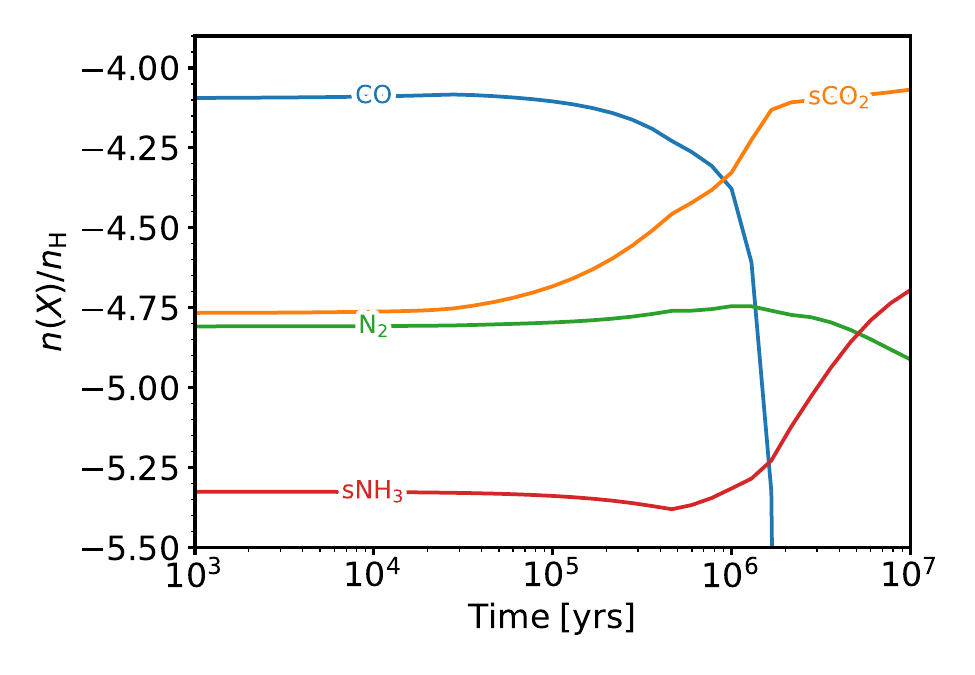}
    \caption{Computed midplane abundance of CO, N$_2$, sCO$_2$ and sNH$_3$ as a function of time and at $r=10$ au. For this figure, salts formation was neglected.}
    \label{fig:one_spot}
\end{figure}

We first illustrate the sink effect affecting the abundance of sNH$_3$ as described above with a model excluding salts formation. Figure \ref{fig:one_spot} shows the computed midplane abundance (i.e. at $z=0$ au) of N$_2$, CO (in the gas phase) as well as sCO$_2$, sNH$_3$ (in the ice form) as a function of time and at $r=10 $ au. 
After $t= 1$ Myr, CO from the gas phase is converted to sCO$_2$ on the surface of grains in a process driven by cosmic rays. The removal of CO from the gas phase allows an efficient conversion of gas-phase N$_2$ into sNH$_3$ at the surface of the grains at $t\gtrsim 1$ Myr. When salts formation is taken into account, most of sNH$_3$ subsequently reacts to form ammonium salts on a timescale $\gtrsim 1$ Myr (see Fig. \ref{fig:fig1}). We note that outside $r \sim 50$ au, nitrogen predominantly exists as sN$_2$ ice on grain surfaces and the conversion process described above is  ineffective in the solid phase. The conversion of CO and N$_2$ into more refractory species at the surface of the grains results in a shift of the radial CO and N$_2$ snowlines closer to the star as a function of time (see also \citet{Ruaud19}): the location of the radial CO snowline shifts from $r\sim 10$ au at $t=1$ Myr to $r\sim 6 $ au at $t=10$ Myrs and the radial N$_2$ snowline shifts from $r\sim 30$ au at $t=1$ Myr to $r\sim 6 $ au at $t=10$ Myrs. It should be noted that at $t=10$ Myrs, the CO and N$_2$ snowlines coincide with each other, even though nitrogen is stored in more refractory species than CO at the surface of the grains (i.e. ammoniated salts). This arises from the fact that the sink effect of N$_2$ is halted by the presence of CO in the gas phase; N$_2$H$^+$ now mostly reacts with CO to re-form N$_2$ and HCO$^+$. This sink effect on gas phase CO and N$_2$ also affects the distribution of N$_2$H$^+$ in the disk midplane which changes as CO and N$_2$ get converted into more refractory species at the surface of grains \citep[see also][]{Ruaud19}.

\subsection{Initial elemental abundance of sulfur}
\label{sec:sulfur} 

As seen in Sec \ref{sec:salt_form}, our model predicts that inside $r\sim 50$ au, most of the sulfur is in \hydrosulfide. However, \hydrosulfide~ only accounts for a small fraction of salts formed at the surface of grains in the fiducial model (i.e. less than $1\%$ with respect to water ice, see Fig. \ref{fig:fig1}).   Recent observations of H$_2$S in comet 67P, on the other hand, point to an abundant reservoir of sulfur at its surface \citep{Altwegg22}. In fact, the simultaneous detection of large quantities of H$_2$S and NH$_3$ after cometary dust impacted ROSINA-DFMS mass spectrometer onboard Rosetta point to the presence of abundant ammonium hydrosulfide (\hydrosulfide) in the comet \citep{Altwegg22}.

The low contribution of \hydrosulfide~ to the total reservoir of salts formed in our model can be explained by the assumed low initial abundance of S in the fiducial model (i.e. S/H=$10^{-6}$). In order to quantify how much sulfur can be stored in the form of salts in the inner disk region, we run an additional disk model where we set the elemental initial abundance of sulfur to $1.5 \times 10^{-5}$ (i.e. its solar value).

\begin{figure} 
    \centering
    \includegraphics[width=1\linewidth]{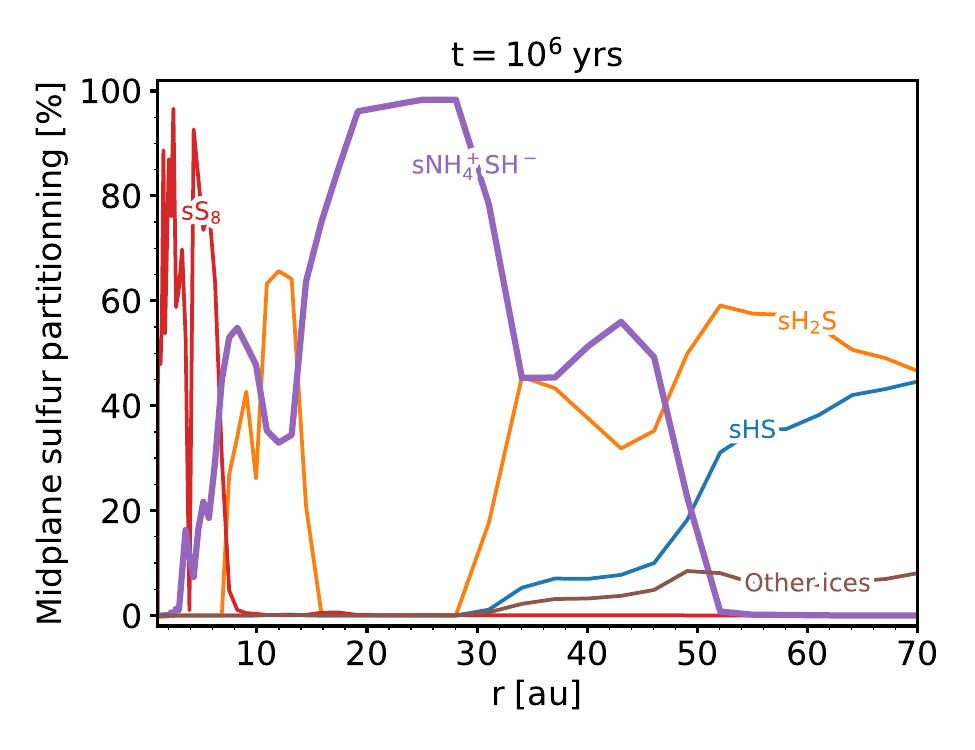}\\
    \includegraphics[width=1\linewidth]{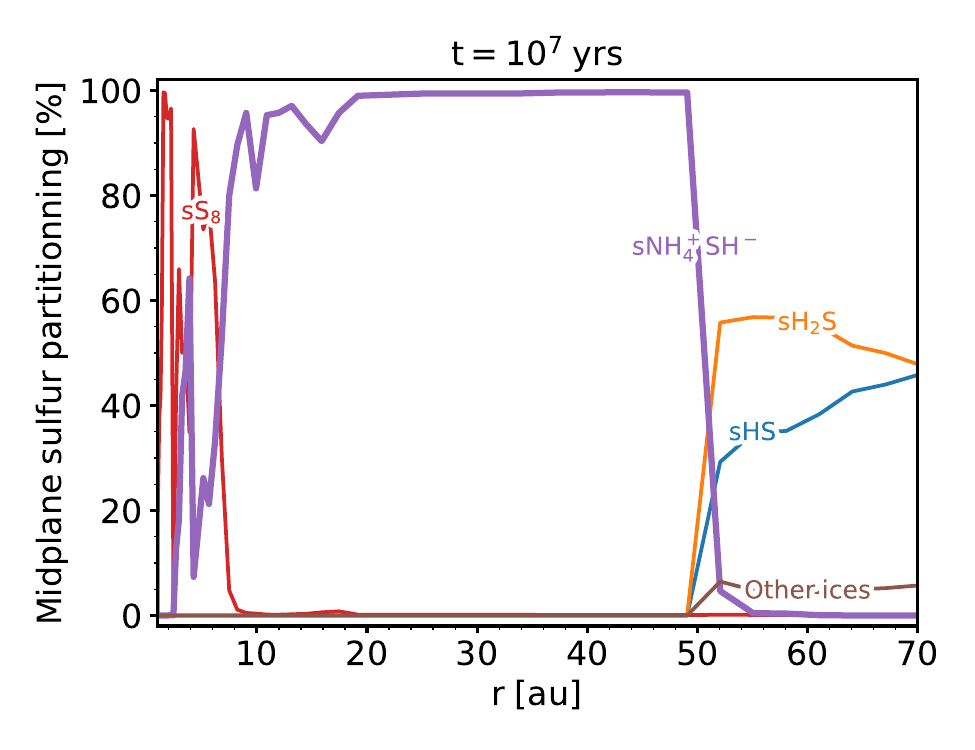}
    \caption{Sulfur partitioning in the disk midplane for the model with an elemental abundance of sulfur of $1.5 \times 10^{-5}$. The top panel is at $t=1$ Myr and the bottom panel at $t=10$ Myrs. We note that there a no major differences in the sulfur partitioning when using an elemental sulfur abundance of $10^{-6}$.}
    \label{fig:sulfur}
\end{figure}

Fig. \ref{fig:sulfur} shows the midplane partitioning of sulfur at $t=1$ Myr and $t=10$ Myrs. From this figure, one can see that ammonium hydrosulfide is predicted to be one of the main carriers of sulfur in the inner disk and that at $t=10$ Myrs, \hydrosulfide~ carries $\sim 100\%$ of the available sulfur in between $r\sim 10$ au and $r\sim 50$ au. As this will be discussed below, increasing the elemental abundance of sulfur to its solar value impacts the midplane partitioning of nitrogen. This is explained by the fact that more N can be stored in the form of \hydrosulfide. Inside $r\sim 8$ au, sS$_8$ is found to be the main reservoir of sulfur in the disk midplane. It forms via successive reactions of sS + sS$_X$, where $X=1$ to 7. This result should, however, be interpreted with caution, as our chemical network remains incomplete with respect to sS$_X$ species. In particular, key formation and destruction pathways for these sulfur-bearing species are not fully accounted for, which may affect their predicted abundances. We emphasize, however, that this limitation primarily impacts the inner disk regions (within $\sim 8$ au), where the sink effect on gas-phase N$_2$ leading to the formation of ammonium salts at the surface of the ice is halted by the presence of CO in the gas-phase (see Sec \ref{sec:sink_N2}). A more comprehensive treatment of sS$_X$ chemistry is deferred to future work.

\subsection{Nitrogen partitioning in the disk midplane}
\label{sec:nitrogen}

\begin{figure*}
    \centering
    \includegraphics[width=1\linewidth]{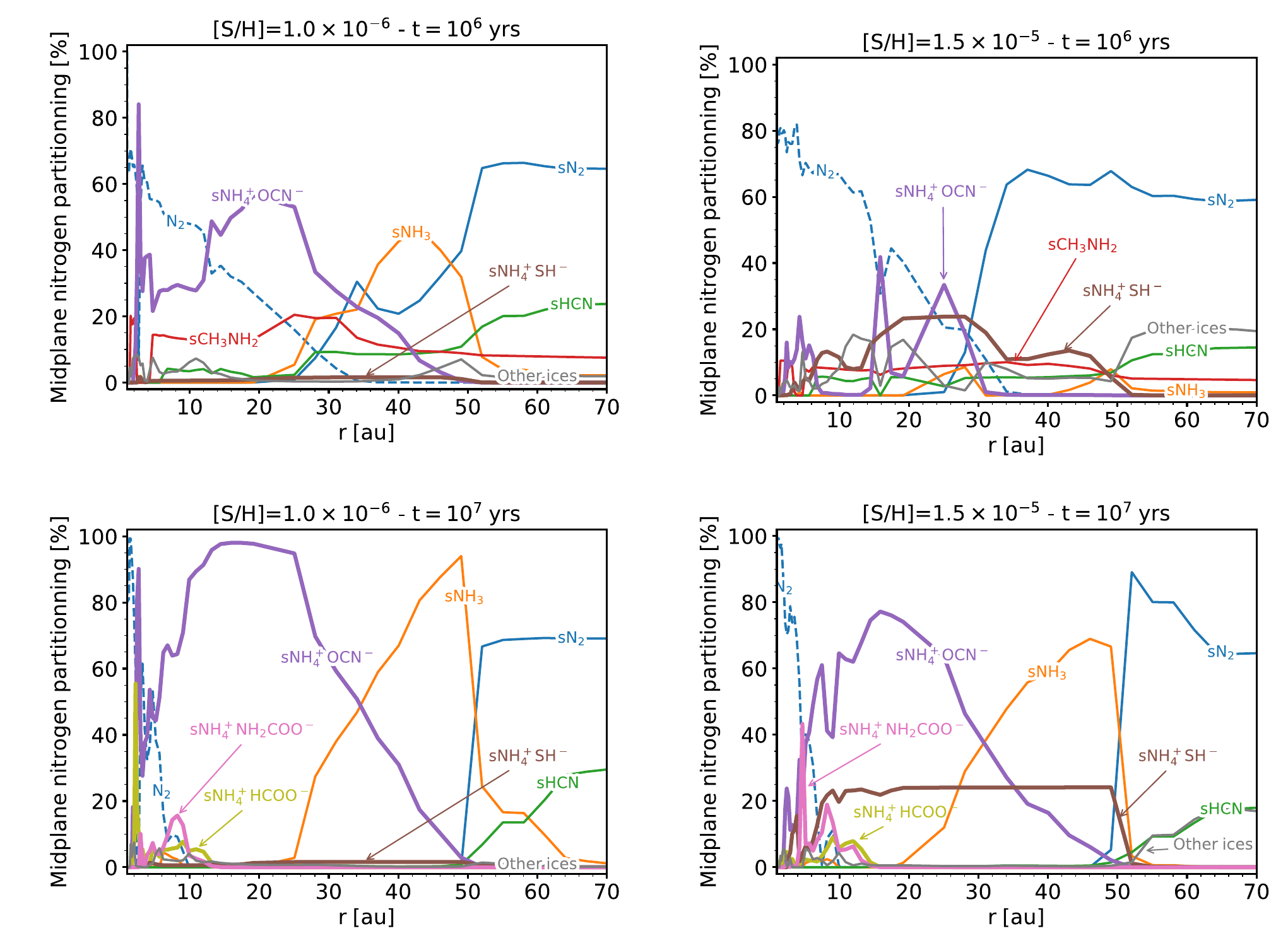}
    \caption{Nitrogen partitioning in the disk midplane for the model with an elemental abundance of sulfur of $10^{-6}$ (left panels) and $1.5 \times 10^{-5}$ (right panels).  Top panels are at $t=1$ Myr and bottom panels at $t=10$ Myrs.}
    \label{fig:fig2}
\end{figure*}

The formation of ammonium salts in the inner disk midplane (i.e. inside $r\sim 50$ au) has an important impact on how nitrogen is distributed within the disk. This is shown in Fig. \ref{fig:fig2}, which depict the nitrogen partitioning in the disk midplane for the model with an elemental abundance of S of $10^{-6}$ and $1.5 \times 10^{-5}$.

For the model with an elemental abundance of S of $10^{-6}$ and at $t=1$ Myr, gas phase N$_2$ is the main nitrogen reservoir inside $r\sim 15$ au followed by ammonium cyanate (sNH$_4^+$OCN$^-$), sCH$_3$NH$_2$ and sHCN. At $r\gtrsim 10$ and $r\lesssim 30$ au, ammonium cyanate becomes the main nitrogen carrier. Outside  of $r\sim30$ au, most of the nitrogen is in sNH$_3$, sN$_2$, sHCN and sCH$_3$NH$_2$. At  $t=10$ Myrs, most of the nitrogen is in the form of ammonium salts at the surface of grains in between $r\sim10$ au and $r\sim25$ au from the star. Ammonium cyanate is, by far, the main nitrogen carrier in the inner disk; i.e. it carries more than $80\%$ of the available nitrogen at $10\lesssim r \lesssim 25$ au.  
This is the result of the cosmic-rays-driven conversion of gas phase N$_2$ to sNH$_3$ and subsequently to salts discussed in Sec. \ref{sec:sink_N2}. We note that dissociation of sCH$_3$NH$_2$, which carries $\sim 20\%$ of the nitrogen in the disk at the beginning of the simulation, also participates in the formation of sNH$_3$. The large predominance of \cyanate in this region compared to other salts is explained by the efficient diffusion of sHCN at the surface of the ice compared to other acidic reactants like sHCOOH which is nevertheless abundant in the ice (typically on the order of few $10^{-5}$ in the inner disk midplane). In between $30 \lesssim r \lesssim 50$ au, most of the nitrogen is in sNH$_3$ while sN$_2$ and sHCN dominate outside $r\sim 50$ au.

The increased abundance of  ammonium hydrosulfide in the model with S/H=$1.5\times 10^{-5}$ compared to the model with S/H=$ 10^{-6}$ discussed in Sec. \ref{sec:sulfur} strongly affects how nitrogen is distributed in the disk midplane. As one can see in the top right panel of Fig. \ref{fig:fig2}, at $t=1$ Myr of the model with S/H=$1.5\times 10^{-5}$, gas phase N$_2$ is still the main nitrogen reservoir inside $\sim 20$ au followed by \cyanate, \hydrosulfide, sCH$_3$NH$_2$ and sHCN. At $t=1$ Myr, \hydrosulfide carries $\sim 10-20\%$ of the total nitrogen in the disk midplane at $r\lesssim 50$ au. At $t=10$ Myrs, \hydrosulfide~ contribution to the nitrogen partitioning reaches a plateau of  $\sim 25\%$ in between $10 \lesssim r\lesssim 50$ au which corresponds to the assumed elemental abundance ratio of S/N$=0.24$ in this model, reflecting the fact that all the sulfur in is \hydrosulfide. The increased abundance of \hydrosulfide~ in this model decreases the amount of nitrogen carried by \cyanate~ inside $r\sim 50$ au by $\sim 20\%$ compared to the model where we assumed S/H=$ 10^{-6}$ (see Fig. \ref{fig:fig2}).

\subsection{Dependence on the cosmic-rays ionization rate}
As seen in previous sections, cosmic rays play a key role in the formation of ammonium salts inside $\sim 30$ au through a chemical sink that converts gas-phase N$_2$ into sNH$_3$ and subsequently to ammonium salts. 
Cosmic ray ionization rates are however relatively uncertain and have a complex dependence both on environment and on column densities \citep[e.g.,][]{Padovani18}. It has also been suggested that these rates could be further reduced in disks by stellar winds and magnetic shielding \citet{Cleeves13}. 
In what follows, we study the effect of decreasing the cosmic-ray ionization rate on the results. We adopt a constant cosmic-ray ionization rate of $\zeta_\mathrm{CR} = 10^{-18}$ s$^{-1}$ throughout the disk, which is the median value inferred from observations of N$_2$H$^+$ in the exoALMA disk sample by \citet{Trapman25} as well as the midplane value computed by \citet{Padovani18} in a magnetized disk around a T Tauri star. This corresponds to a decrease of $\sim 1$ order of magnitude compared to the midplane value obtained from the depth dependent expression of \citet{Padovani18} we use in the fiducial disk model.

\begin{figure}
    \centering
    \includegraphics[width=1\linewidth]{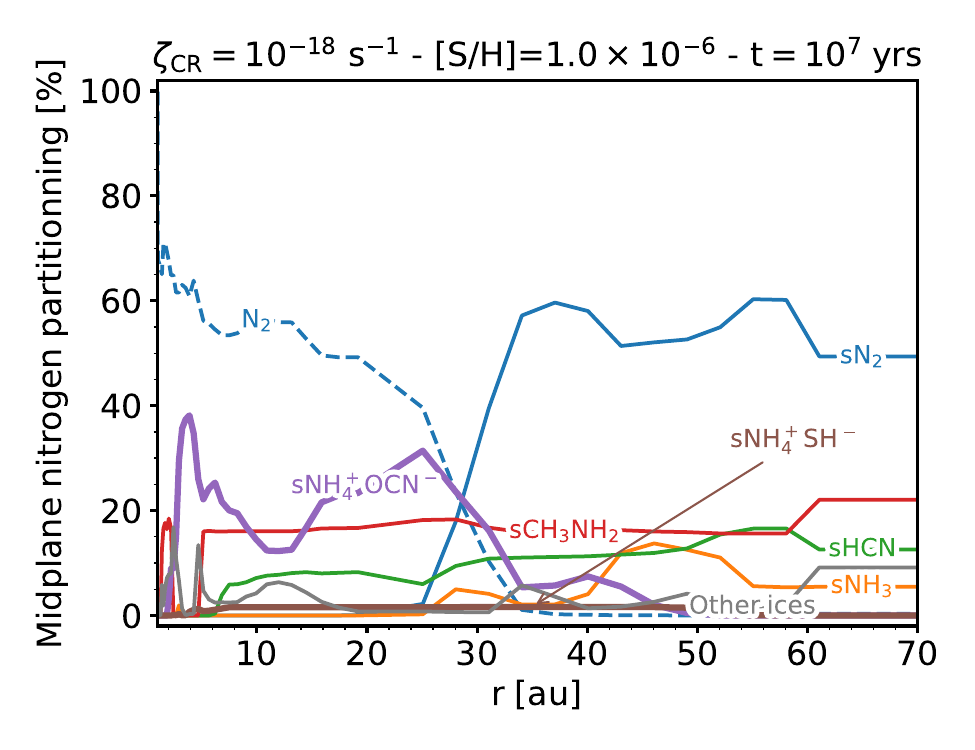}
    \caption{Nitrogen partitioning in the disk midplane computed at $t=10$ Myrs for the model with an elemental abundance of sulfur of $10^{-6}$ and a low cosmic-ray ionization rate of $\zeta_\mathrm{CR}=10^{-18}$s$^{-1}$.}
    \label{fig:fig3}
\end{figure}

Fig. \ref{fig:fig3} shows the midplane partitioning of nitrogen in the disk midplane computed at $t=10$ Myrs for a model with an elemental sulfur abundance of $10^{-6}$ and $\zeta_\mathrm{CR} = 10^{-18}$ s$^{-1}$. As seen in this figure, the amount of nitrogen stored in the form of ammonium salts decreases compared to the fiducial model (bottom left panel of Fig. \ref{fig:fig2}). Inside $r\sim 30$ au, \cyanate, which is the dominant salt in the ice, accounts for $\sim 20-40 \%$ of the total available nitrogen. This corresponds to a decrease of a factor $\sim 2-5$ compared to the fiducial model, and explained by the fact that the timescale for the formation of ammonium salts by the sink effect on gas-phase N$_2$ increases when decreasing $\zeta_\mathrm{CR}$.

\section{Discussion} 
\label{sec:discussion}

In previous sections, we showed that ammonium salts form efficiently in the inner disk midplane (i.e. inside $r\sim 50$ au, which corresponds to the comet-forming region of the solar nebula) and that they dominate the nitrogen and sulfur budget of the ice after 1 Myr. In particular, we showed that inside the N$_2$ snowline, gas-phase N$_2$ is efficiently and irreversibly removed by a cosmic-ray-driven chemical sink that converts volatile nitrogen into sNH$_3$ and subsequently to ammonium salts in the ice on Myr timescales.

\subsection{Potential observational diagnostics in disks}
This work shows that ammonium salts may constitute a major reservoir of sulfur and nitrogen in protoplanetary disks. Among them, \cyanate, predicted by our models to be the most abundant ammonium salt, may indeed be present in disks, as suggested by the detection of NH$_4^+$ and OCN$^-$ in two objects with JWST \citep{Sturm23,Potapov25} and  OCN$^-$ in more recent JWST disk observations \citep{Bergner2026}. \hydrosulfide, predicted to be the second most abundant salt, could also represent a major nitrogen and sulfur reservoir. Detecting SH$^-$ in disk ices with JWST would provide additional constraints on (1) the presence of ammonium salt in disks, and (2) the existence of a significant sulfur reservoir. Interpreting JWST ice observations, however, remains challenging. Absorption bands often result from the overlap of multiple molecular species, making it difficult to unambiguously identify the carrier of a given feature. Nevertheless, a few spectral regions accessible to JWST may enable the detection of SH$^-$. In particular, the broad $\mathrm{NH}_4^+$ $v_4$ + $\mathrm{SH}^-$ libration combination band at 5.3 $\mu$m appears to be a promising tracer of SH$^-$ in the icy disk midplane, owing to the absence of other major ice features at this wavelength, even though it might be weak \citep{Slavicinska25}. Another potential observational diagnostic of salt is through inward dust drift, which can carry ice-coated dust grains into the inner disk, where sublimation at the corresponding snowlines enhances their gas-phase abundance \citep[e.g.][]{Booth19,Nakazawa26}. In the case of salts, this could result in an enhanced abundance of their dissociated products in the inner disk midplane, i.e. at $ 1\lesssim r\lesssim 3$ au, where $ 120 \lesssim T \lesssim 200$ K, corresponding to the salt desorption temperature range. It is interesting to note that recent ALMA observations in the disk around HD163296 have revealed the existence of warm H$_2$S emission that most likely originates from the inner disk regions (i.e. $r\sim$ 3 au) \citet{Yamato26}. Lastly, a possible observational diagnostic of salt is through N$_2$H$^+$. Our models predict that the sink effect on gas phase CO and N$_2$ affects the distribution of N$_2$H$^+$ in the disk midplane (i.e. at $z/r \lesssim 0.15$). However,  including salt formation has only a minor effect on the computed N$_2$H$^+$ emission, as most of the emission arises from regions beyond $r \sim 50$ au, where salt formation is inefficient. Even for the fraction of N$_2$H$^+$ emission originating inside $r \sim 50$ au, the impact of salt formation remains negligible. This is because the emission mainly originates from regions near the vertical water and CO snowlines (i.e. at $z/r \sim 0.15$). In this region, the chemistry is dominated by photoprocessing of the ice by stellar and interstellar photons and the sink effect on gas phase N$_2$ is reduced.

\subsection{Link with comets}
When comparing with observations in comets, our model predicts that the abundance of salts -- and ammonium cyanate in particular -- relative to water ice can reach $\sim 10\%$ to $30\%$ at $t\gtrsim 1$ Myr, which is $\sim 1$ order of magnitude larger than nitrogen observations in cometary coma (i.e. assuming that most of  nitrogen volatiles in comets form by dissociation of ammonium salts in the coma), where NH$_3$/H$_2$O $\sim 0.2-5 \%$ \citep[e.g.][]{Biver24}. On the other hand, based on the study of the reflectance spectra of 67P, \citet{Poch20} estimate that almost $\sim 50\%$ of the total nitrogen budget (including refractory material) could be in ammonium salts with only $\sim 1\%$ in volatile species and the remaining into more refractory material, bringing the total cometary N close to the solar value. If such a large abundance of ammonium salts is indeed present at the surface of 67P, this could indicate that the abundance of species like NH$_3$, HNCO and HCOOH found in cometary comae might not reflect the abundance of their associated salts on cometary nucleus. Ammonium salts thermally desorb at higher temperatures than that of water ice, and contrary to many other species, salts seem not to co-desorb with water. In the case of ammonium cyanide and ammonium formate mixed with water ice, \citet{Noble13} and \citet{Kruczkiewicz21} find that these salts remain on the surface after the complete desorption of the water. In line with these results, in comet 67P, typical fragments of salts were only detected after cometary dust impacted ROSINA-DFMS mass spectrometer onboard Rosetta \citep{Altwegg20,Altwegg22}. The abundance of NH$_3$ with respect to H$_2$O increased by a factor of $\gtrsim 100$ compared to before the impact, reaching NH$_3$/H$_2$O $\sim 1$. A similar behavior was also observed for H$_2$S \citep{Altwegg22}. Consequently, observations of water and possible salt fragments in the cometary coma cannot reliably constrain their relative abundances in nucleus ices, as current evidence in 67P suggests that salts may be confined to cometary dust. 

\subsection{Limitations}
Finally, our results should be regarded as upper limits on the abundance of ammonium salts in disks, as we did not account for any destruction processes in the ice other than desorption. In particular, \citet{Vitorino24} show that exposing \hydrosulfide~ to hydrogen atoms is effective in reducing its amount, although with a reduced efficiency compared to other hydrogenation reactions. This could be explained by the fact that part of the salt reforms after hydrogenation \citep{Vitorino24}. To conclude, although partial data exist on the formation of ammonium salts in ices, their reactivity remains largely unexplored, despite their importance  protoplanetary disks and in icy bodies of planetary systems. In this study, the formation of the most abundant salt, \cyanate, is proposed to proceed via the reaction of another salt (sO + \cyanade), whose reactivity has not been thoroughly investigated, either theoretically or experimentally. This highlights the need for further study. In fact, these salts may not be inert end products but active intermediates in ice chemistry.

\section{Conclusion} 
\label{sec:conclusion}

In this paper we study the formation of ammonium salts in protoplanetary disks. We use a disk modeling framework that self-consistently computes the disk structure by imposing vertical hydrostatic equilibrium and solves for time-dependent gas-grain chemistry. The results of this work can be summarized as follows:
\begin{enumerate} 
    \item Ammonium salts form efficiently in the inner disk midplane ($r \lesssim 50$ au), where thermal diffusion enables their constituent species to meet and react on icy grain surfaces. Beyond $\sim 50$ au, grains are too cold for diffusion, thereby suppressing salt formation. \cyanate~ and \hydrosulfide~ readily form throughout the inner disk midplane via surface diffusion of sHCN, sHNCO, and sH$_2$S, whereas \formate, \acetate, and \carbamate~ are confined to within $\sim 15$ au, where higher temperatures facilitate reactions between precursors with larger diffusion energy barriers.
    \item Inside $r \sim 30$ au, ammonium salt formation is enhanced by a cosmic-ray-driven sink that progressively converts gas-phase CO and N$_2$ into sCO$_2$ and sNH$_3$ ices on timescales $\gtrsim 1$ Myr. 
    \item Inside $r \sim 50$ au, nitrogen and sulfur are predominantly in the form of ammonium salts on icy grain surfaces when using our fiducial depth-dependent cosmic-ray ionization rate. \cyanate~ is found to be the dominant nitrogen reservoir in the disk midplane over $10 \lesssim r \lesssim 30$ au, accounting for $\gtrsim 80\%$ of the total nitrogen budget at $t = 10$ Myr. Adopting an elemental sulfur abundance of $1.5 \times 10^{-5}$, \hydrosulfide~ becomes the second most abundant salt beyond $\sim 10$ au, incorporating $\sim 20\%$ of the nitrogen and $\sim 100\%$ of the sulfur at $t = 10$ Myr.
    \item Ammonium salt formation in the inner regions of protoplanetary disks depends on the adopted cosmic-rays ionization rate. A one order of magnitude decrease in the cosmic-rays ionization rate (i.e., going from $\zeta_{CR}\sim 10^{-17}$ s$^{-1}$ to $\zeta_{CR}\sim 10^{-18}$ s$^{-1}$) results in a decrease by a factor of $\sim 2-5$ in the amount of nitrogen stored in the form of salts at the surface of the grains inside $r\sim 50$ au.
\end{enumerate} 
 
JWST may provide additional constraints on the presence of salts to the already detected features associated with \cyanate~ (i.e. NH$_4^+$ and OCN$^-$) in protoplanetary disks. However, interpreting JWST observations is challenging, making the unambiguous identification of salt signatures difficult. Nevertheless, a few JWST-accessible spectral regions may contain additional features attributable to the presence of ammonium salts in disks.

\section*{Acknowledgments}
We are thankful to the anonymous referee who provided
comments that improved this contribution. U.G. acknowledges
support from the NASA Interdisciplinary Consortia for Astrobiology Research (Grant No. NNH22ZDA001N-ICAR).

\bibliography{sample701}{}
\bibliographystyle{aasjournalv7}

\appendix
\section{NH$_2$COOH}
\label{sec:NH2COOH}

Table \ref{tab:tab3} contains all bimolecular reactions added to the network for the formation and destruction of sNH$_2$COOH ice. Based on our model, sNH$_2$COOH mainly forms in the inner disk midplane (i.e. at $r\lesssim 50$ au) where molecules and radicals can efficiently diffuse across grain surfaces due to higher temperatures. In this region, it is formed by two main reaction routes. The first one is given by
\begin{equation}
    \mathrm{sNH_2} + \mathrm{sHOCO} \rightarrow \mathrm{sNH_2COOH}
\end{equation}
were sHOCO is formed by sOH + sCO.
The second formation route is through
\begin{equation}
    \mathrm{sOH} + \mathrm{sNH_2CO} \rightarrow \mathrm{sNH_2COOH},
\end{equation}
where sNH$_2$CO is mainly formed by 
\begin{equation}
\mathrm{sCN} + \mathrm{sH_2O} \rightarrow \mathrm{sNH_2CO}.
\end{equation}
Inside $r\sim 15 $ au, most of sNH$_2$COOH is destroyed to form \carbamate~ by the reaction of
\begin{equation}
    \mathrm{sNH_3} + \mathrm{sNH_2COOH} \rightarrow \mathrm{sNH_4^+NH_2COOH^-}
\end{equation}
Outside $r\sim 15 $ au, photodissociation by cosmic-rays photons dominate sNH$_2$COOH destruction.

\begin{table*}
    \centering
\begin{tblr}{lllcccX}
        \hline
        \hline
         \SetCell[c=3]{c}Reaction  & & & $\Delta E$& Branching & $E_a$ & Notes \\
         & & & [kJ/mol] & Ratio & [K] & \\
         \hline
         $\mathrm{sH  } + \mathrm{sHNCO  }$ &  $\rightarrow$ & $\mathrm{sNH_2CO   } $ & -105 & 1.0 & 4900 & \SetCell[r=2]{l} M06-2X/AVTZ calculations (this work) in good agreement with \citet{Nguyen1996}.\\
         \\
         $\mathrm{sH  } + \mathrm{sNH_2CO }$ &  $\rightarrow$ & $\mathrm{sNH_2CHO }$ & -384 & 0.6 & 0 & \SetCell[r=4]{l} The first steps are likely the barrierless NH2CHO formation and also direct H atom abstraction. The TS for NH$_2$CHO $\rightarrow$ NH$_3$ + CO is located 351 kJ/mol above the NH$_2$CHO so some, but likely few, of the NH$_2$CHO should not dissociate into NH$_3$ + CO. \citet{Rimola2018} study suggest NH$_2$CHO and H$_2$ + HNCO production only.\\
								   &  $\rightarrow$ & $\mathrm{sNH_3     } +  \mathrm{sCO   }$ & -342 & 0.2 & 0 \\
								   &  $\rightarrow$ & $\mathrm{sHNCO    } +  \mathrm{sH_2   }$ & -322 & 0.2 &0 \\
        \\
        $\mathrm{sH  } + \mathrm{sNH_2CHO}$ &  $\rightarrow$ & $\mathrm{sH_2      } +  \mathrm{sNH_2CO}$ & -42 & 1.0 & 3400 & M06-2X/AVTZ calculations (this work). See also \citet{Haupa2019} \\
        $\mathrm{sNH } + \mathrm{sHCO   }$ &  $\rightarrow$ & $\mathrm{sHNCO    } +  \mathrm{sH  }$ & -308 & 0.1 & 0 & \SetCell[r=3]{l} The first steps are likely the barrierless NHCHO formation and also direct H atom abstraction. NHCHO will quickly evolve toward HNCO + H through a low exit TS and toward NH$_2$CO and then NH$_2$ + CO. \\
								  &  $\rightarrow$ & $\mathrm{sNH_2     } +  \mathrm{sCO }$ & -318 & 0.9 & 0 \\	
        \\
        $\mathrm{sNH } + \mathrm{sH_2CO  }$ &  $\rightarrow$ & $\mathrm{sNH_2CHO  }          $ &-436 & 0.2 & 3000  & \SetCell[r=3]{l} The first step is the abstraction of a hydrogen atom through a calculated barrier of 4300K at the M06-2X/AVTZ level, which decreases significantly in the presence of water molecules. The resulting NH$_2$ and HCO then react to form either NH$_2$CHO or NH$_3$ + CO.\\
								   &  $\rightarrow$ & $\mathrm{sNH_3     } +  \mathrm{sCO}    $ & -397 & 0.8 & 3000 \\
        \\
        $\mathrm{sNH_2} + \mathrm{sCO    }$ &  $\rightarrow$ & $\mathrm{sNH_2CO   }          $ & -95 & 1.0 & 2100  & \SetCell[r=1]{l} M06-2X/AVTZ calculations (this work)\\
        $\mathrm{sNH_2} + \mathrm{sHCO} $ &  $\rightarrow$ & $ \mathrm{sNH_2CHO} $ &-419 & 0.2 & 0 & \SetCell[r=3]{l} The first steps are likely the barrierless NH$_2$CHO formation and also direct H atom abstraction. The TS for NH$_2$CHO $\rightarrow$ NH$_3$ + CO is located 351 kJ/mol above the NH$_2$CHO. Some H$_2$ + HNCO may also be produced. See also \citet{Rimola2018}\\
        								  &  $\rightarrow$ & $ \mathrm{sNH_3   } + \mathrm{sCO} $ & -380 & 0.8 & 0 \\
        \\
        $\mathrm{sNH_2} + \mathrm{sHOCO  }$ &  $\rightarrow$ & $\mathrm{sNH_2COOH }          $ & -441 & 0.5 & 0 & \SetCell[r=3]{l} The first steps are likely the barrierless NH$_2$COOH formation and also direct H atom abstraction. The TS from NH$_2$COOH $\rightarrow$ NH$_3$ + CO$_2$ is located +179 kJ/mol above NH$_2$COOH, so some of the NH$_2$COOH will dissociate.\\
								     &  $\rightarrow$ & $\mathrm{sNH_3     } +  \mathrm{sCO_2}$ & -433 & 0.5 & 0\\
        \\
        $\mathrm{sNH_3} + \mathrm{sCO    }$ &  $\rightarrow$ & $\mathrm{sNH_2CHO  }          $ & -39 & 1.0 & 36000 & \SetCell[r=1]{l} M06-2X/AVTZ calculations (this work)\\
        $\mathrm{sNH_3} + \mathrm{sCO_2   }$ &  $\rightarrow$ & $\mathrm{sNH_2COOH }          $ & -7 &1.0 & 21000 & \SetCell[r=1]{l} M06-2X/AVTZ calculations (this work)\\
        $\mathrm{sCN} + \mathrm{sH_2O   }$ &  $\rightarrow$ & $\mathrm{sNH_2CO }             $ &-225 & 1.0 & 0 & \SetCell[r=3]{l} In the gas phase, the only possible pathway is the one producing HCN + OH through a barrier equal to 2900K at M06-2X/AVTZ (3500K from \citet{Jacobs1989}). In ice, a second mechanism is possible, involving several water molecules and producing NH$_2$CO through submerged barriers, which are likely globally barrierless according to \citet{Rimola2018}.\\
                                    &  $\rightarrow$ & $\mathrm{sHCN } + \mathrm{sOH }             $ &-37 & 0.0 & 3000 \\
        \\
        \hline 
\end{tblr}
\caption{ Bimolecular reactions added for the formation $\mathrm{sNH_2COOH}$, which is not part of KIDA. We also added reactions for the formation of $\mathrm{sNH_2CO}$.}
    \label{tab:tab3}
\end{table*}

\begin{table*}
    \centering
\begin{tblr}{lllcccX}
        \hline
        \hline
         \SetCell[c=3]{c}Reaction  & & & $\Delta E$& Branching & $E_a$ & Notes \\
         & & & [kJ/mol] & Ratio & [K] & \\
         \hline
        $\mathrm{sO  } + \mathrm{sNH_2CHO}$ &  $\rightarrow$ & $\mathrm{sNH_2COOH }          $ & -496 & 0.1 & 900  & \SetCell[r=4]{l} The first step is the abstraction of a hydrogen atom through a calculated barrier of 900K at the M06-2X/AVTZ level. OH will further reacts with NH$_2$CO leading mainly to NH$_2$COOH and HNCO + H$_2$O. The TS from NH$_2$COOH $\rightarrow$ NH$_3$ + CO$_2$ is located +179 kJ/mol above NH$_2$COOH, so some of the NH$_2$COOH will dissociate.\\
								     &  $\rightarrow$ & $\mathrm{sNH_3     } +  \mathrm{sCO_2}$ & -488 & 0.4 & 900 \\
								     &  $\rightarrow$ & $\mathrm{sHNCO    } +  \mathrm{sH_2O}$ & -420 & 0.5 & 900 \\
        \\
        $\mathrm{sOH } + \mathrm{sNH_2CO }$ &  $\rightarrow$ & $\mathrm{sHNCO    } +  \mathrm{sH_2O}$ & -381 & 0.3 & 0 & \SetCell[r=4]{l} The first steps are likely the barrierless NH$_2$COOH formation and also direct H atom abstraction. The TS from NH$_2$COOH $\rightarrow$ NH$_3$ + CO$_2$ is located +179 kJ/mol above NH$_2$COOH, so some of the NH$_2$COOH will dissociate into NH$_3$ + CO$_2$.
 \\
								     &  $\rightarrow$ & $\mathrm{sNH_2COOH }          $ & -457 & 0.1 & 0 \\
								     &  $\rightarrow$ & $\mathrm{sNH_3     } +  \mathrm{sCO_2}$ & -450 &0.6 & 0 \\
        \\
        $\mathrm{sOH } + \mathrm{sNH_2CHO}$ &  $\rightarrow$ & $\mathrm{sNH_2CO   } +  \mathrm{sH_2O}$ & -102 &1.0 &  0 & \SetCell[r=2]{l} M06-2X/AVTZ leads to a submerged barrier in the gas phase (-6 kJ/mol). See also \citet{Bunkan2016}.\\
        \\
        \hline 
\end{tblr}
\caption{Continuation of Table \ref{tab:tab3}}
\end{table*}

\end{document}